\documentclass[preprint,aps,showpacs,nofootinbib]{revtex4}

\usepackage{epsfig,amssymb,amsmath}

\newcommand{\beq}{\begin{equation}}
\newcommand{\eeq}{\end{equation}}
\newcommand{\bea}{\begin{eqnarray}}
\newcommand{\eea}{\end{eqnarray}}
\newcommand{\bear}{\begin{array}}
\newcommand {\eear}{\end{array}}
\newcommand{\bef}{\begin{figure}}
\newcommand {\eef}{\end{figure}}
\newcommand{\bec}{\begin{center}}
\newcommand {\eec}{\end{center}}
\newcommand{\non}{\nonumber}

\newcommand{\la}{\left\langle}
\newcommand{\ra}{\right\rangle}

\def\EQ#1{Eq.~(\ref{#1})}

\def\REF#1{(\ref{#1})}

\def\lrf#1#2{ \left(\frac{#1}{#2}\right)}
\def\lrfp#1#2#3{ \left(\frac{#1}{#2} \right)^{#3}}

\begin{document}
\draft
\tighten
\preprint{TU-928}
\title{\large \bf
The Spectral Index and its Running in Axionic Curvaton
}
\author{
      Fuminobu Takahashi\footnote{email: fumi@tuhep.phys.tohoku.ac.jp}}
\affiliation{
 Department of Physics, Tohoku University, Sendai 980-8578, Japan
    }

\vspace{2cm}

\begin{abstract}
We show that a sizable running spectral index suggested by the recent SPT data 
can be explained in the axionic curvaton model with a potential that consists of two 
sinusoidal contributions of different height and period. 
We find that the running spectral index is generically given by $d n_s/d \ln k \sim \frac{2\pi}{\Delta N} (n_s - 1)$,
where $\Delta N$ is the e-folds during one period of modulations. 
In the string axiverse,   axions naturally acquire a mass from multiple contributions, and one of the axions
may be responsible for the density perturbations with a sizable running spectral
index via the curvaton mechanism.  We note that the axionic curvaton model with modulations 
can also accommodate the red-tilted spectrum with a negligible running, without relying on large-field inflation.
\end{abstract}

\maketitle


\section{Introduction}
The origin of density perturbations is one of the central issues in cosmology.
While the quantum fluctuation of the inflaton is the simplest possibility,
there may be many light scalars in nature, which acquire quantum fluctuations 
extending beyond the horizon during inflation. If so, some of them may give a significant contribution to the
observed density perturbations via the curvaton~\cite{Linde:1996gt,Enqvist:2001zp,Lyth:2001nq,Moroi:2001ct} 
(or its variant, e.g. modulated reheating~\cite{Dvali:2003em,Kofman:2003nx}) mechanism.
Those models can be constrained or preferred by observations, especially if one finds an
extension(s) of the LCDM cosmology such as non-Gaussianity, running spectral index, dark radiation,
isocurvature perturbations, and so on.

Interestingly, the recent SPT data gives preference to a large negative running spectral index~\cite{Hou:2012xq}\footnote{
On the other hand, the ACT data~\cite{Sievers:2013wk} does not give preference for any extensions of the LCDM model.
The SPT and ACT observations look at different patches of the sky,
and hopefully the tension will be resolved by the Planck or future observations.
}
\beq
\frac{d n_s}{d \ln k}\;=\; -0.024 \pm 0.011
\eeq
for the CMB  (SPT+WMAP7) data alone. The significance increases further if the BAO and $H_0$ data are included. 
In a single-field slow-roll inflation models with a featureless potential, the running spectral index is second
order in the slow-roll parameters, and therefore it is of order $(n_s-1)^2 \sim10^{-3}$. Thus, it is a challenge to explain
the observed running, if taken at face value. There have been various proposals. See e.g.~Refs.~\cite{Chung:2003iu,Cline:2006db,Espinosa:2006pb,Joy:2007na,Joy:2008qd,Kawasaki:2003zv,Yamaguchi:2003fp,Easther:2006tv,Kobayashi:2010pz}.

%
%

In this letter we consider a curvaton scenario~\cite{Linde:1996gt,Enqvist:2001zp,Lyth:2001nq,Moroi:2001ct} 
 in order to explain the observed running spectral index. 
We consider an axionic curvaton model in which a pseudo Nambu-Goldstone boson 
plays a role of curvaton. In order to explain the running, we assume that the curvaton
potential  consists of two sinusoidal contributions of different height and period. 
It is possible to generate such potential, if the axion mass receives contributions from several
instantons,  which may be ubiquitous in the string axiverse~\cite{Arvanitaki:2009fg}.
As we shall see later, the running spectral index $\alpha \equiv d n_s/d\ln k$ is generically 
given by
\beq
\alpha \;\sim \; \frac{2\pi}{\Delta N}\, (n_s - 1),
\eeq
where $\Delta N$ is the e-folding number during one period of the modulations. 
Thus, the observed  spectral index and its running can be
naturally explained if $\Delta N = {\cal O}(10)$.

The rest of this letter is organized as follows. In Sec.~\ref{sec:2}, we summarize the latest observational results
and the prediction of the spectral index and its running in the curvaton scenario. We present our model in Sec.~\ref{sec:3}
and show that the observed running can be explained in the axionic curvaton model with repeated modulations.
The last section is devoted for discussion and conclusions.

\section{Spectral index and its running}
\label{sec:2}
In this section we summarize the latest SPT result and the expression of the
spectral index $n_s$ and its running $\alpha$ in terms of the curvaton potential $V(\sigma)$, which
will provide us with implications for the curvaton model building. 

The scale dependence of primordial curvature perturbations can be parametrized by
the spectral index $n_s$ and its running $\alpha$ defined by~\cite{Kosowsky:1995aa}
\bea
n_s(k) &=& \frac{d \ln \Delta_{\cal R}^2(k)}{d \ln k},\\
\alpha(k)     &=&  \frac{d n_s(k)}{d \ln k},
\eea
where $\Delta_{\cal R}^2(k)$ represents  a power spectrum of primordial curvature perturbations ${\cal R}_k$,
and $k$ is a comoving wave number. 
For a constant running, the power spectrum is given in the following form,
\bea
\Delta_{\cal R}^2(k) & = & \Delta_{\cal R}^2(k_0) \lrfp{k}{k_0}{n_s(k_0)-1+ (\alpha/2) \ln \lrf{k}{k_0}}.
\eea
The latest SPT data combined with the WMAP 7yr data give preference to the non-zero
running at more than $2 \sigma$~\cite{Hou:2012xq}
\bea
\alpha &=& -0.024 \pm 0.011
\eea
with $n_s(k_0^{\rm SPT}) \approx 0.96$ at the SPT pivot scale $k_{0}^{\rm SPT} = 0.025\,{\rm Mpc}^{-1}$.
If we extrapolate the above result to the WMAP pivot scale $k_{0}^{\rm WMAP} = 0.002\,{\rm Mpc}^{-1}$,
the spectral index is roughly estimated to be $n_s(k_{0}^{\rm WMAP}) \approx 1.02 > 1$.  Thus, the spectral
index changes from blue-tilted to red-tilted as one goes from  $k_{0}^{\rm WMAP}$ to  $k_{0}^{\rm SPT}$.
The e-folding number between the
two pivot scales is $\Delta N_{\rm pivot} = \ln (k_{0}^{\rm SPT}/k_{0}^{\rm WMAP}) \simeq 2.5$.

In the curvaton scenario, the spectral index and its running can be expressed in terms of the
curvaton potential $V(\sigma)$ and the inflation scale~\cite{Kobayashi:2012ba}:
\bea
\label{nscurv}
n_s & \simeq & 1+ 2 \frac{\dot{H}_*}{H_*^2}+ \frac{2 V''(\sigma_*)}{3 H_*^2}, \\
\label{alphacurv}
\alpha & \simeq &2 \frac{\ddot{H}_*}{H_*^3}-4 \frac{\dot{H}_*^2}{H_*^4} - \frac{4}{3}
\frac{\dot{H}_*}{H_*^2} \frac{V''(\sigma_*)}{H_*^2} - \frac{2}{9} \frac{V'(\sigma_*) V'''(\sigma_*)}{ H_*^4},
\eea
where $\sigma_*$ and $H_*$ are the curvaton field value and the Hubble parameter
when the cosmological scale $k$ exited the horizon during inflation. The prime denotes the derivative
with respect to $\sigma$. 

Let us make a simplifying assumption that
the Hubble parameter hardly evolves during inflation, which is justified except for the large-field inflation.\footnote{ 
We note that the following argument can be applied to large-field inflation in a straightforward way.
} Then, the above expressions can be simplified as
\bea
\label{ns}
n_s & \simeq & 1+ \frac{2 V''(\sigma_*)}{3 H_*^2}, \\
\label{alpha}
\alpha & \simeq & - \frac{2}{9} \frac{V'(\sigma_*) V'''(\sigma_*)}{ H_*^4}.
\eea
In order for $n_s$ to change from $n_s > 1$ to $n_s < 1$, therefore, the curvature of the potential must change its sign
from positive to negative as the curvaton evolves. This requires some structure in the curvaton potential.
However, if such structure is present only in a limited field range, we would encounter another problem:
why cosmological scales exited the horizon just when the curvaton passed the region where the structure exists. 
This fine-tuning problem can be avoided if the potential has a structure
everywhere in the potential. Indeed, Kobayashi and the present author applied this
idea to the large-field inflation by adding repeated modulations to the inflaton potential, and
showed that the large negative running
can be realized without affecting the overall inflaton dynamics~\cite{Kobayashi:2010pz}.
This should be contrasted to the result of Ref.~\cite{Easther:2006tv} that the total e-folds of inflation does
not exceed about $30$ in the presence of large negative running.
In the same spirit of Ref.~\cite{Kobayashi:2010pz}, we consider the curvaton potential with modulations in the next section.

\section{Axionic Curvaton}
\label{sec:3}
We consider an axionic curvaton model, in which the curvaton is a pseudo-Nambu-Goldstone boson.
The shift symmetry is assumed to be explicitly broken by e.g. non-perturbative effects as in the QCD
axion, generating the following potential:
\beq
V_0(\sigma)\;=\; \Lambda_1^4 \left(1-\cos\lrf{\sigma}{f_1} \right).
\label{v0}
\eeq
In the following we will focus on $-\pi < \sigma/f_1 < \pi$ without loss of generality.
The axionic curvaton has been studied in e.g. Refs.~\cite{Dimopoulos:2003az,Kawasaki:2008mc,
Kawasaki:2011pd,Kawasaki:2012gg}.  The model has an advantage that the curvaton mass
is protected by the (approximate) shift symmetry which enables the curvaton to acquire quantum fluctuations
during inflation.\footnote{It is possible to generate density perturbations with a red-tilted spectral index and
 a vanishingly small running,  if $\sigma$ initially sits in the vicinity of
the hilltop, without relying on large-field inflation. The hilltop initial condition
generically predicts a sizable non-Gaussianity $f_{\rm NL} = {\cal O}(10)$. 
This enhancement of the non-Gaussianity is due to non-uniform onset of curvaton oscillations~\cite{Kawasaki:2008mc,Kawasaki:2011pd}.
The importance of non-uniform onset of oscillations in the curvaton scenario was first pointed out in Ref.~\cite{Kawasaki:2008mc}
to our knowledge. 
}

The curvaton $\sigma$ acquires quantum fluctuations $\delta \sigma = H_{\rm inf}/2\pi$ during inflation, which
turn into density perturbations after the curvaton dominates the Universe. Unless the initial position is close to the
hilltop, the WMAP normalization of density perturbations~\cite{Hinshaw:2012fq} imposes the following relation,
\beq
\lrfp{H_{\rm inf}}{2 \pi \sigma_*}{2} \;\simeq\; 2.4 \times 10^{-9},
\label{wmap}
\eeq
where $H_{\rm inf}$ is the Hubble parameter during inflation.\footnote{
Precisely speaking, it is necessary to take account of the non-uniform onset of oscillations. 
However it was shown in~\cite{Kawasaki:2008mc,
Kawasaki:2011pd,Kawasaki:2012gg} that the predicted density perturbations are close to that in the case of quadratic potential
unless the initial position is  close to the hilltop.
}  Thus, for a given inflation scale $H_{\rm inf}$,
 the WMAP normalization determines $\sigma_*$, whose typical value is $f_1$.
Here and in what follows we assume that the curvaton
dominates the Universe at the decay. We shall come back to this issue in Sec.~\ref{sec:4}.

As we have seen in the previous section, the observed value of the running spectral index indicates that the spectral index is blue at the WMAP pivot 
scale, and it is red at the SPT pivot scale. Thus, the curvature of the potential must be
initially positive and gradually become negative as the curvaton evolves.   Furthermore, the deviation of $n_s$ from unity
should be comparable to its running. 

First let us see that  it is impossible with
 the curvaton potential \REF{v0}  to generate such sizable running. 
Using \REF{ns}, \REF{alpha} and \REF{v0},  we obtain
\bea
n_s - 1 &\simeq & \frac{2}{3} \frac{\Lambda_1^4 }{ f_1^2 H_*^2} \cos \lrf{\sigma_*}{f_1},\\
\alpha  &\simeq &\frac{2}{9} \frac{\Lambda_1^8 }{ f_1^4 H_*^4} \sin^2 \lrf{\sigma_*}{f_1}.
\label{alpha_single}
\eea
We note that the spectral index changes from red ($n_s < 1$) to blue ($n_s > 1$), but
not the other way around. Thus, the running can be only positive, which is also manifest from \EQ{alpha_single}. 
In addition, 
the running is generically of order $(n_s - 1)^2$, therefore
it is too small to account for the observed value~\cite{Hou:2012xq}.
This necessitates some modifications to the curvaton potential \REF{v0}. 

\begin{figure}[t]
\begin{center}
\includegraphics[scale=0.8]{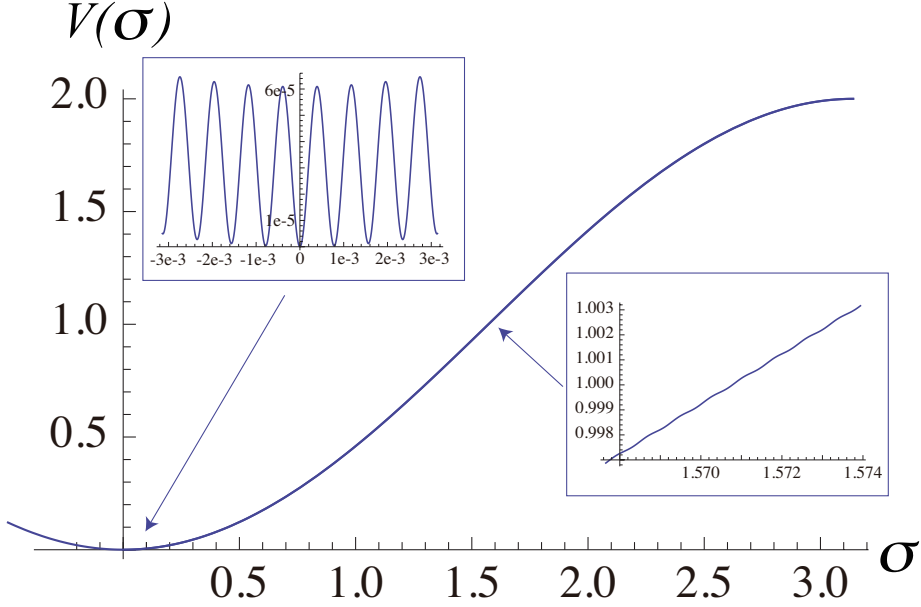}
\caption{
The potential $V(\sigma)$ is shown. We set
$m_1 = 0.01 H_{*}$, $m_2 = 0.44 H_*$, and
$f_2 = 1.25 \times 10^{-4} f_1$, and
the potential and the curvaton field
are normalized by $m_1^2 f_1^2$ and $f_1$, respectively.  We can see the modulations $\delta V$
is subdominant around the inflection point, but it becomes significant around the origin. 
}
\label{fig1}
\end{center}
\end{figure}

Let us add  modulations $\delta V(\sigma)$ to the original potential $V_0(\sigma)$:
\bea
V(\sigma)&=& V_0(\sigma) + \delta V(\sigma),
\label{v0dv}
\eea
with
\beq
\delta V(\sigma) \;=\;\Lambda_2^4\, \left(1-\cos\left(\frac{\sigma}{f_2} + \theta \right)\right).
\eeq
The relative phase $\theta$ is not relevant for the following analysis, and so, we will set $\theta=0$.
See Fig.~\ref{fig1}. 
Such two contributions can be originated from the axion anomalous couplings with two different
gauge fields, and the multiple contributions to the axion mass are considered to be ubiquitous in the string axiverse~\cite{Arvanitaki:2009fg}. 
Alternatively we may consider a complex scalar field $\Phi$ with an approximate
global U(1) symmetry broken by non-renormalizable operators.
For instance, consider a scalar potential,
\beq
V(\Phi) \;=\; - m_\Phi^2 |\Phi|^2 + \frac{|\Phi|^{2n-2}}{M_*^{2n-6}} 
+\left( a_n \frac{\Phi^n}{M_*^{n-3}}  + a_m \frac{\Phi^m}{M_*^{m-3}} + {\rm h.c.}\right)
\eeq
where  $M_*$ is a cut-off scale and we assume $n < m$. Then $\Phi$ develops a vev at 
$\la \Phi \ra\sim (m_\Phi M_*^{n-3})^{1/(n-2)}$ and the U(1) symmetry gets spontaneously broken.
 The phase of $\Phi$ is a pseudo-Nambu-Goldstone boson
and receives two sinusoidal potentials of different height and period from the breaking terms in the parenthesis. 
Such a scalar potential can be realized in a supersymmetric theory with a discrete (R) symmetry. 

For convenience let us introduce two mass scales, $m_i = \Lambda_i^2/f_i$, where $i$ runs over $1,2$. 
We assume the following hierarchy among the model parameters~\cite{Kobayashi:2010pz},
\bea
\label{cond1}
m_1^2 f_1^2 &\gg& m_2^2 f_2^2,\\
\label{cond2}
m_1^2 f_1 &\gg& m_2^2 f_2,\\
\label{cond3}
m_1^2  &\ll& m_2^2,
\eea
which imply $f_1 \gg f_2$.
The first two conditions ensure the curvaton dynamics is basically determined by $V_0$ for most of the field
range except for the vicinity of the extrema of $V_0$. On the other hand, the third condition tells us that
the local curvature of the potential is dictated by the modulations, $\delta V$.\footnote{Apart from the running spectral index, such a set-up will be
useful when we would like to realize a red-tilted spectral index in curvaton scenario without relying on
large-field inflation which necessitates super-Planckian variation of the inflaton. The modulation generates a red-tilted spectrum, while it is $V_0$
that determines the curvaton abundance. } In particular, the WMAP normalization condition \REF{wmap} is maintained,
while the spectral index and its running can be significantly affected by the modulations $\delta V$.

Let us now derive conditions on the modulations $\delta V$ to realize the observed spectral index and the running. 
From \REF{nscurv} and \REF{cond3}, the curvature  of $\delta V$ should satisfy
\beq
m_2^2 \cos\lrf{\sigma_*}{f_2} \;\simeq\; \pm\, {\cal O}(0.01)\, H_{\rm inf}^2
\eeq
in order to realize $n_s - 1 = \pm\,{\cal O}(0.01)$ at the SPT and WMAP pivot scales. 
On the other hand, the slow-roll condition reads 
\beq
m_2^2 \;\lesssim\; {\cal O}(0.1)\, H_{\rm inf}^2.
\eeq
The above two conditions imply
\bea
m_2^2 &\simeq&{\cal O}(0.1)\, H_{\rm inf}^2,\\
\cos\lrf{\sigma_*}{f_2} & \simeq & \pm\, {\cal O}(0.1).
\eea
Furthermore, since the spectral index changes its sign between the two pivot scales, 
the curvaton should vary by about $c f_2$ with $c = 0.1 \sim 1$ during the e-folds
$\Delta N_{\rm pivot} \simeq 2.5$.

Let us evaluate the variation of the curvaton $\Delta \sigma$ while the two pivot scales exited the horizon during inflation.
The equation of motion can be approximated as
\beq
3 H_{\rm inf} \dot{\sigma} \;\simeq\; - m_1^2 f_1 \sin \lrf{\sigma}{f_1},
\label{eom}
\eeq
where the dot denotes the derivative with respect to time. 
Since we are interested in the variation of $|\Delta \sigma| \lesssim f_2 \ll f_1$, we may
approximate the rhs of \REF{eom} to be a constant. Thus,  $\Delta \sigma$
is related to the increase of the e-folding number $\Delta N_{\rm pivot}$ as
\beq
\Delta \sigma \;\simeq\; - \frac{ m_1^2 f_1}{3 H_{\rm inf}^2}  \sin \lrf{\sigma_*}{f_1}\Delta N_{\rm pivot} .
\label{deltas}
\eeq
Combined with $\Delta \sigma = c f_2$, we obtain
\beq
\frac{ m_1^2 f_1}{3 H_{\rm inf}^2}  \sin \lrf{\sigma_*}{f_1} \Delta N_{\rm pivot} \;=\; c f_2
\label{period}
\eeq
with $c \approx 0.1 - 1$.

The decay constant $f_2$  determines the period of the modulations. Both the spectral index and its running
return to its original value as the curvaton varies by $2\pi f_2$. (Here we have neglected the change of $V_0$ during one period,
because $f_1 \gg f_2$.)  As the  SPT (and WMAP) data implies an almost constant running over the observed CMB
scales, the curvaton should not vary by more than
the half of the period $\pi f_2$ while the CMB scales exited the horizon. The corresponding e-folding number
is roughly $\Delta N_{CMB} \simeq \ln(\ell_{\rm max}) \simeq 8$, where $\ell_{\rm max} \approx 3000$ for the SPT data.
Using \REF{deltas}, this constraint reads
\beq
\frac{ m_1^2 f_1}{3 H_{\rm inf}^2}  \sin \lrf{\sigma_*}{f_1} \Delta N_{\rm CMB} \;< \; \pi f_2,
\label{cmb}
\eeq
which can be satisfied for $f_2$ given by \REF{period}.

\begin{figure}[t]
\begin{center}
\includegraphics[scale=0.8]{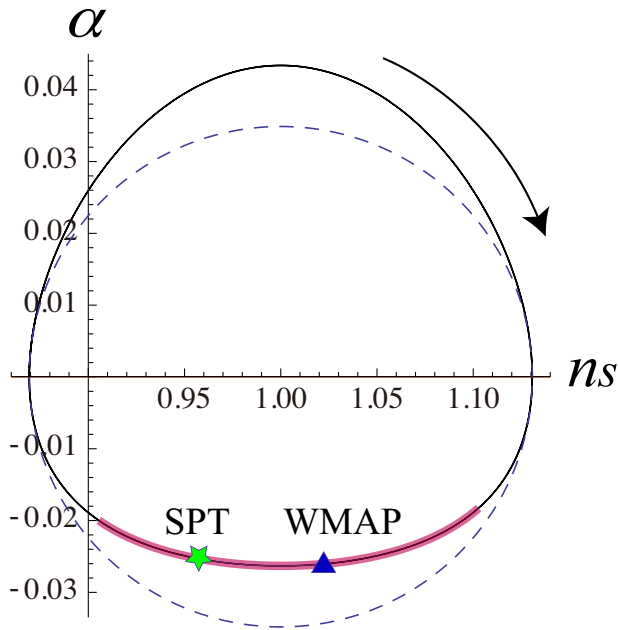}
\caption{
The evolution of the spectral index $n_s$ and the running $\alpha$. 
The numerical result is shown by the solid line,
while the dashed line represents the analytic solution \REF{ana_ns} and \REF{alc}.
 As the curvaton evolves,
the predicted $(n_s, \alpha)$ moves clockwise along an oval as indicated by the arrow, and it rotates 
twice during about $50$ e-folds. We also show
the SPT and WMAP pivot scales by the star and the triangle, respectively. The CMB scales $(\ell \lesssim 3000)$
are shown as the thick line.  
}
\label{fig2}
\end{center}
\end{figure}

Let us write the spectral index and the running in terms of the curvaton potential.
The spectral index is
\bea
n_s-1 &\simeq& \frac{2}{3} \frac{m_2^2}{ H_*^2} \cos\lrf{\sigma_*}{f_2},
\label{ana_ns}
\eea
and its running is given by
\bea
\alpha &\simeq & \frac{2}{9} \frac{m_1^2 f_1 m_2^2}{f_2 H_*^4} \sin\lrf{\sigma_*}{f_1} \sin\lrf{\sigma_*}{f_2}.
\label{alc}
\eea
Thus we can see that both $n_s$ and $\alpha$ oscillate due to the modulations as the curvaoton evolves.
In particular,  it moves clockwise along an oval in the $(n_s, \alpha)$-plane, as shown in Fig.~\ref{fig2}.
The oscillation amplitudes $A_{n_s}$ and $A_\alpha$ are related as
\beq
A_{\alpha} \;\simeq\; \frac{c}{\Delta N_{\rm pivot}} A_{n_s},
\eeq
where we have used \EQ{period}.

Similarly, we can easily show that the oscillation amplitude of the running spectral index is generally related to 
that of $(n_s-1)$ as 
\beq
A_\alpha \;\simeq\; \frac{2\pi}{\Delta N}  A_{n_s},
\eeq
where $\Delta N$ denotes the e-folds during one period of the modulations $\delta V$. 
Schematically we may express this result as
\setlength{\fboxrule}{1pt}
\beq
\boxed{
\hspace{3mm}
\alpha \sim \frac{2\pi}{\Delta N} (n_s-1)
\hspace{3mm}
}
\label{alphans}
\eeq
where $\alpha$ and $n_s - 1$ are understood to represent their typical values.
The observed spectral index and the running can be explained for $\Delta N = 20 \sim 30$.
(The condition \REF{cmb} requires $\Delta N/2 > \Delta N_{\rm CMB} \sim 8$.)

We note that there is an upper bound on $\Delta N$ as
\bea
\Delta N &=& 2 \pi f_2 \left(\frac{m_1^2 f_1}{3 H_{\rm inf}^2} \sin \lrf{\sigma_*}{f_1} \right)^{-1},\non\\
& < & 
 \frac{4 \pi}{|n_s-1|}   \frac{|\cos \lrf{\sigma_*}{f_2}|}{\sin \lrf{\sigma_*}{f_1}}.
 \label{updn}
\eea
Therefore, for e.g. $n_s \simeq 0.97$, $\Delta N$ is smaller than a few hundred.  In other words,
$\alpha$ is always larger than of order$(n_s-1)^2$ in our framework. 

%
%
%

Lastly let us present the numerical results. We set $m_1 = 0.01 H_{*}$, $m_2 = 0.44 H_*$, and
$f_2 = 1.25 \times 10^{-4} f_1$. The WMAP normalization \REF{wmap} will determine the relation between
$f_1$ and $H_*$, and so, all the mass scales are determined if one fixes $f_1$ (or $H_*$). The following numerical
results do not depend on specific values of $f_1$. The only assumption is that the curvaton dominates
the Universe before the decay. This condition depends on the thermal history of the Universe, especially
the reheating temperature, as well as the decay rate and the oscillation amplitude of the curvaton. We will discuss this issue in the
next section.

In Fig.~\ref{fig1}, we show the potential $V(\sigma) = V_0(\sigma) + \delta V(\sigma)$. We can see that
the modulations do not change much the slope of the potential around the inflection point, while
they are significant around the origin. The evolution of the curvaton after the commencement of oscillations
are considered to be more complicated than without modulations, which will be discussed in Sec.\ref{sec:4}.

The evolution of $n_s$ and $\alpha$ is shown in Fig.~\ref{fig2}. As the curvaton evolves,
the predicted $(n_s, \alpha)$ moves clockwise along an oval as indicated by the arrow, and it rotates 
twice during about $50$ e-folds, i.e., $\Delta N \simeq 25$.  The solid line represents the numerical result, while the
the dashed line represents the analytic solution \REF{ana_ns} and \REF{alc}.
We also show
the SPT and WMAP pivot scales by the star and the triangle, respectively. The CMB scales $(\ell \lesssim 3000)$
are shown as the thick line. The position of $\sigma_*$ is set to be around the inflection point of
$V_0$, but the result is not sensitive to the value of $\cos(\sigma_*/f_1)$ unless it is extremely close to the
extrema of $V_0$. (Note that the value of $\cos(\sigma_*/f_2)$ determines the position of the CMB scales along
the orbit.) 
The reason why the actual orbit is not a circle is that our analysis relies on the
approximation \REF{cond1} - \REF{cond3}. In particular,  $\delta V'$ partially cancels (enhances)
$V_0'$ when the $\alpha$ is negative (positive). This reduces the change of the running if it is negative.
We can see from Fig~\ref{fig2} that the orbit is slightly flattened at the bottom and
 the running is more or less constant over the CMB scales.
We can see that the observed spectral index and the running can be
indeed realized in our model. 

%
%

\section{Discussion and conclusions}
\label{sec:4}

Let us here discuss the evolution of the curvaton after inflation. When the Hubble parameter  $H$
becomes comparable to $m_1$, the curvaton starts to oscillate with an amplitude of order $f_1$. The modulations $\delta V$
does not affect this timing even though the curvature of the potential is determined by $\delta V$.
This was also confirmed by numerical calculations. As the Universe expands, the oscillation energy
evolves as non-relativistic matter.  The situation changes
when the oscillation energy becomes comparable to the height of the modulations $\delta V \sim m_2^2 f_2^2$. 
The oscillation amplitude at the time is about $(m_2/m_1) f_2$, which is much larger than $f_2$, and so, there are
many local minima within the oscillation amplitude. 
Then the curvaton will be trapped by one of those minima, 
and the oscillation (angular) frequency suddenly changes from $m_1$ to $m_2$. 

One may expect that domain walls are formed during this transition, because the curvaton will be trapped
by different vacuum at different places\footnote{We expect that
the curvaton fluctuations with a characteristic frequency are produced through parametric resonance
 while the curvaton oscillates in a potential with modulations. Then the curvaton oscillations may be spatially
 inhomogeneous at subhorizon scales. 
}
Actually however the domain wall formation is hindered by the
bias, i.e., the difference in the energy density  among those minima, which arises from $V_0$. 
To see this,  let us first assume the inflaton-matter dominated Universe. 
Then the transition takes place when $H_f \sim m_2 (f_2/f_1)$. 
The bias $\epsilon$ between the adjacent vacua will be minimized in the vicinity of the 
origin as
\beq
\epsilon \;\gtrsim\; m_1^2 f_2^2.
\eeq
If the domain walls are formed, the tension is given by $\sim m_2 f_2^2$.
Assuming that the scaling law is achieved soon after the formation, the domain walls annihilate effectively
when the force of tension acting on the domain walls becomes comparable to the pressure arising from
the bias, i.e., $H_{\rm ann} \sim m_1 (m_1/m_2)$. However we can see that
\bea
H_f & > & H_{\rm ann} 
~\Longrightarrow~ m_2^2 f_2  \;\gtrsim\;  m_1^2 f_1,
\eea
which contradicts with \REF{cond2}. This means that the domain walls are not formed from the beginning because
of the large bias in the inflaton-matter domination era. On the other hand, if the reheating is completed at the commencement
of oscillations, the above condition is modified as
\bea
H_f &>& H_{\rm ann} 
~\Longrightarrow~ m_2^2 f_2^\frac{2}{5}  \;\gtrsim\;  m_1^2 f_1^\frac{2}{5},
\eea
which may be satisfied depending on the parameters. For the parameters adopted in the numerical calculation,
this condition is marginally satisfied, and so,  the domain walls will annihilate quickly even if they formed. The domain
walls may play an important role for another choice of parameters, especially in a more general context of the
moduli problem in the string axiverse. We leave this for future work.  In any case the domain walls are not likely formed,
or even if they are formed, they will soon annihilate in the present case. 

After the reheating of the inflaton, the fraction of the curvaton energy density will increases until it decays.
In order for the curvaton mechanism to work, the curvaton energy should be sizable at the curvaton decay. 
The fraction of the curvaton energy should be greater than $\sim 0.01$ in order not to generate too large
non-Gaussianity. This often requires a large inflation scale as well as high reheating temperature. 
Let us comment on how to realize the curvaton domination. For a fixed inflation scale,
the decay constant $f_1$ is more or less fixed unless we adopt the hilltop initial condition. 
A smaller $m_1$ would delay the commencement of oscillations, but in the inflaton-matter domination era, 
it does not increase the energy fraction of the curvaton.\footnote{The fraction of the curvaton energy density
increases in the radiation domination for a smaller $m_1$.
}
Thus, it may be necessary to reduce the curvaton decay rate $\Gamma_\sigma$.  If the curvaton has an anomalous
coupling to massless gauge bosons with a strength of $1/f_i$, the decay rate is considered to be $\Gamma_\sigma \sim
(N_g \alpha^2/256\pi^2) \,m_2^3/f_i^2$, where $\alpha$ denotes the gauge coupling constant, and $N_g$ denotes the
number of decay modes. The decay rate will be suppressed 
if the gauge coupling is sufficiently small. Alternatively it is possible that the curvaton decays only into
a pair of light fermions through the following interaction,
\beq
{\cal L} \;\supset\; m_\psi\, e^{i \sigma/f_i} \bar{\psi} \psi,
\eeq
where such coupling arises if $\psi_{L(R)}$ is charged under the shift symmetry. 
Then the decay rate is suppressed by the light fermion mass as $\Gamma_\sigma \sim m_\psi^2 m_2/f_i^2$,
which will help the curvaton to dominate the Universe.

Here we comment on the difference from Ref.~\cite{Kobayashi:2010pz}. First, we have considered a 
curvaton model, while the large-field inflation was assumed in \cite{Kobayashi:2010pz}. Therefore,
the tensor mode tends to be smaller in the present case, although the present scenario can be straightforwardly applied
to large-field inflation,  which produces the tensor mode within the reach of future experiments. 
 A relation between $\alpha$ and $n_s$ like \REF{alphans}
holds in the case of  large-field inflation with modulations, if the correction due to the time evolution of $H$ is taken into
account.  On the other hand, it is possible to generate large non-Gaussianity in our model 
if the fraction of the curvaton energy is of ${\cal O}(0.01-0.1)$ at the decay. 
One important difference is the ease of model building; for the present scenario to work,
we need only two sinusoidal potential of different height and period, which can be easily realized in a usual field theory,
and no super-Planckian variation of the scalar field is required.

So far we have focused on the generation of density perturbations with a large running spectral index. In fact we can use the model \REF{v0dv}
to generate a red-tilted spectrum $n_s \sim 0.97$ with a negligible running, without relying on large-field
inflation.\footnote{
Note that the running $\alpha$ is still larger than of order $(n_s-1)^2$, because of \REF{updn}. 
} If there is no structure such as modulations, the red-tilted spectrum implies a sizable (tachyonic) mass of the curvaton.
Thus, the curvaton will start to oscillate soon after inflation ends unless it initially sits in the vicinity of the hilltop~\cite{Kawasaki:2011pd}. 
However, this conclusion can be avoided if we add modulations to the curvaton potential. The red-tilted spectrum $1-n_s = {\cal O}(0.01)$
with a negligible running can be realized if we consider large $\Delta N$ (i.e., large $f_2$) with $m_2^2/H_*^2 = {\cal O}(0.01)$.
This is because the curvaton dynamics (the onset of oscillations as well as the energy density) is determined by $V_0$,
while it is the modulations that affect the spectral index and its running. 

In this letter we have shown that the sizable running spectral index can be explained in the axionic
curvaton model in which the curvaton potential receives two sinusoidal contributions of differrent
height and period. The spectral index and the running are generally related to as \EQ{alphans},
and they are comparable in size for $\Delta N = 20 \sim 30$. Such axions with a potential that consists of 
multiple contributions may be ubiquitous in the string axiverse.

\section*{Acknowledgment}
The author is grateful to Takeshi Kobayashi for fruitful discussion in the early stage of this
work and for reading the manuscript. The author also thanks Tetsutaro Higaki for useful comments. 
This work was supported by the Grant-in-Aid for Scientific Research on Innovative
Areas (No.24111702, No. 21111006, and No.23104008), Scientific Research (A)
(No. 22244030 and No.21244033), and JSPS Grant-in-Aid for Young Scientists (B) 
(No. 24740135) [FT]. This work was also supported by World Premier International Center Initiative (WPI
Program), MEXT, Japan

\end{document}